\documentclass{nature}

\usepackage{braket}
\usepackage{graphicx}
\usepackage{amsmath}

\newcommand{\ex}{\textbf{e}_x}
\newcommand{\es}{\textbf{e}_s}

\title{Imaging topology of Hofstadter ribbons}

\author{Dina Genkina$^{1}$, Lauren M. Aycock$^{1,2,3}$,  Hsin-I Lu$^{1,4}$, Mingwu Lu$^{1}$, Alina M. Pineiro$^{1}$,  \& I. B. Spielman$^{1}$}

\begin{document}

\maketitle

\begin{affiliations}
 \item Joint Quantum Institute, National Institute of Standards and Technology, and University of Maryland, Gaithersburg, Maryland, 20899, USA
 \item Department of Physics, Cornell University, Ithaca, NY 14850
 \item Currently APS/AAAS Congressional Science Fellow
 \item Currently Modern Electron, Bellevue, WA 98007
\end{affiliations}

\begin{abstract}
Physical systems with non-trivial topological order find direct applications in metrology\cite{Klitzing1980} and promise future applications in quantum computing\cite{Freedman2001,Kitaev2003}. The quantum Hall effect derives from transverse conductance, quantized to unprecedented precision in accordance with the system's topology\cite{Laughlin1981}. At magnetic fields beyond the reach of current condensed matter experiment, around $\mathbf{10^4}$ Tesla, this conductance remains precisely quantized but takes on different values\cite{Thouless1982}. Hitherto, quantized conductance has only been measured in extended 2-D systems. Here, we engineered and experimentally studied narrow 2-D ribbons, just 3 or 5 sites wide along one direction, using ultracold neutral atoms where such large magnetic fields can be engineered\cite{Jaksch2003,Miyake2013,Aidelsburger2013,Celi2014,Stuhl2015,Mancini2015}. We microscopically imaged the transverse spatial motion underlying the quantized Hall effect. Our measurements identify the topological Chern numbers with typical uncertainty of $\mathbf{5\%}$, and show that although band topology is only properly defined in infinite systems, its signatures are striking even in nearly vanishingly thin systems.
\end{abstract}

	The importance of topology in physical systems is famously evidenced by the quantum Hall effect's role as an ultra-precise realization of the von Klitzing constant $R_K = h/e^2$ of resistance\cite{Klitzing1980}. Although topological order is only strictly defined for infinite systems, the bulk properties of macroscopic topological systems closely resemble those of the corresponding infinite system. For 2-D systems in a magnetic field $B_0$, the topology is characterized by an integer invariant called the Chern number. Even at laboratory fields of tens of Tesla, crystalline materials have a small magnetic flux $\Phi=AB_0$ per individual lattice plaquette (with area $A$) compared to the flux quantum $\Phi_0=h/e$. Superlattice \cite{Geisler2004,Melinte2004,Feil2007,Dean2013} and ultracold atom\cite{Miyake2013,Aidelsburger2013} systems now realize 2-D lattices in a regime where the magnetic flux per plaquette $\Phi$ is a significant fraction of $\Phi_0$. 

	Experimental signatures of Chern numbers generally leverage one of two physical effects: in condensed matter systems the edge-bulk correspondence allows the Chern number to be inferred from the quantized Hall conductivity $\sigma_{\rm{H}} = C/R_{\rm{K}}$, and in cold-atom experiments direct probes of the underlying band structure give access to the Chern number\cite{Jotzu2014,Aidelsburger2015,Wang2013}.  Both of these connections derive from the pioneering work of  Thouless, Kohmoto, Nightingale, and den Nijs\cite{Thouless1982}, in the now famous TKNN paper.  Going beyond these well known techniques, the TKNN paper showed that for rational flux $\Phi/\Phi_0 = P/Q$ (for relatively prime integers $P$ and $Q$) the integer solutions  $s$ and $C$ to the Diophantine equation 
\begin{equation}
1 = Q s - P C
\label{eqn:Diophantine}
\end{equation}  
uniquely\footnote{Subject to the constraint $|C|\leq |Q|/2$\cite{Thouless1982, Kohmoto1989}. The integer $s$ has no physical significance as changing the flux by a multiple of $2\pi$ simply changes $s$ by a multiple of $C$.} determine the Chern number $C$. We leveraged this result to determine the Chern number of our system.

\begin{figure}
\includegraphics[scale=0.9]{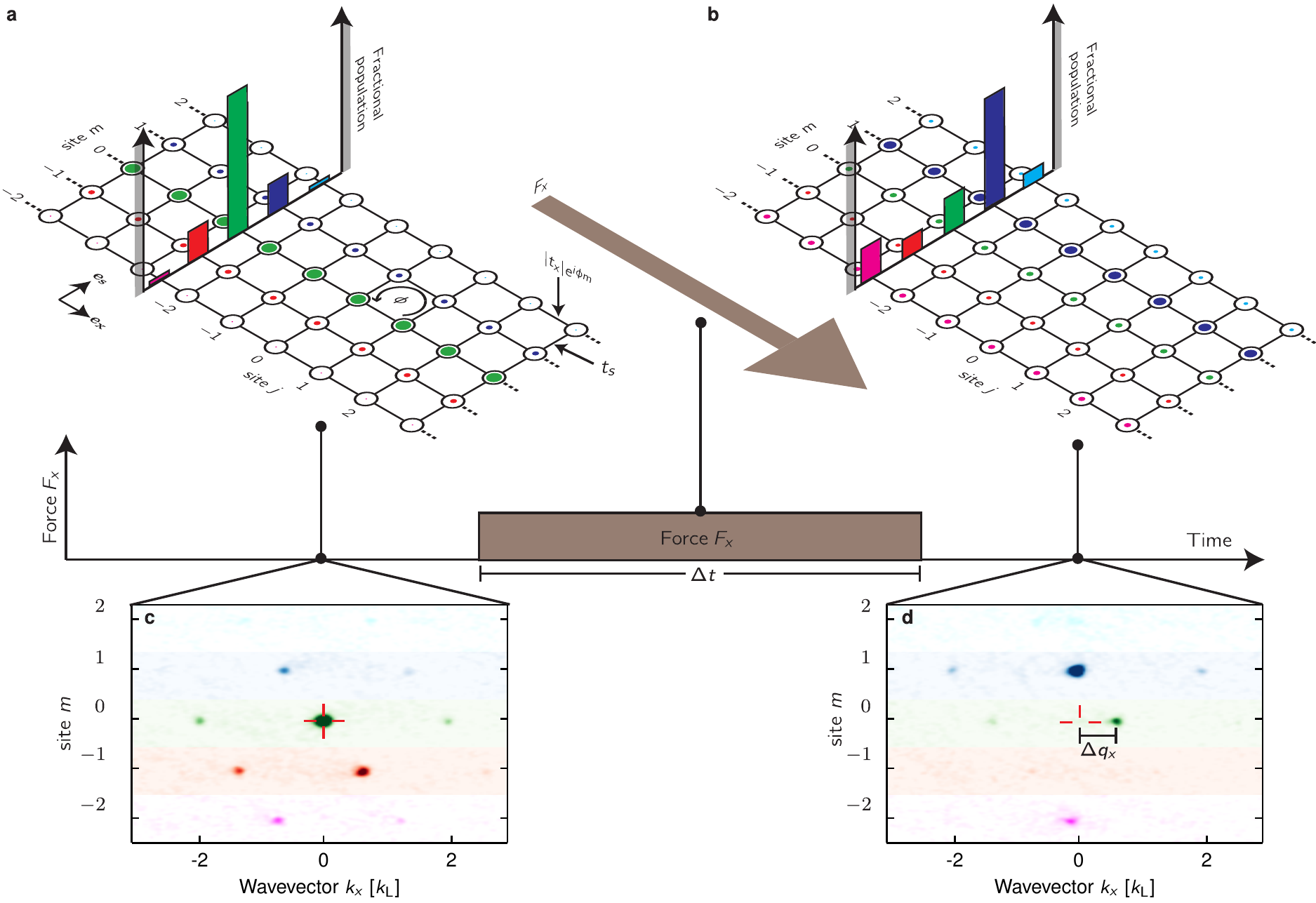}
\caption{Quantum Hall effect in Hofstadter ribbons. \textbf{a}. 5-site wide ribbon with real tunneling coefficients along $\es$   and complex tunneling coefficients along $\ex$,  creating a non-zero phase  $\phi$ around each plaquette.  \textbf{b}. After applying a force along $\ex$  for a time $\Delta t$, atomic populations shift transversely along $\es$, signaling the Hall effect. \textbf{c,d}.  TOF absorption images giving hybrid momentum/position density distributions $n(k_x,m)$. Prior to applying the force \textbf{c}, the $m=0$ momentum peak is at $k_x=0$, marked by the red cross. Then, in \textbf{d}, the force directly changed $q_x$, evidenced by the displacement $\Delta q_x$ of crystal momentum, and via the Hall effect shifted population along $\es$. }
\label{fig:laughlinPump}
\end{figure}
	
	We studied ultracold neutral atoms in a square lattice with a large magnetic flux per plaquette. As pictured in  Fig. \ref{fig:laughlinPump}a, our system consisted of a 2-D lattice that is extremely narrow along one direction, just 3 or 5 sites wide - out of reach of traditional condensed matter experiments, with hard wall boundary conditions: a ribbon. Our system was well described by the Harper-Hofstadter Hamiltonian\cite{Harper1955,Hofstadter1976}
\begin{equation}
\hat{H}= -\sum\limits_{m,j}\left({|t_x| e^{i\phi m}|j,m\rangle\langle j+1,m|+t_s|j,m\rangle\langle j,m+1|}\right) + \rm{H.c.},
\label{eqn:Hamiltonian}
\end{equation}

where $j$ and $m$ label lattice sites along $\ex$    and $\es$, with tunneling strengths $t_x$ and $t_s$ respectively. As shown in Fig. \ref{fig:laughlinPump}a, tunneling along $\ex$    was accompanied by a phase shift $e^{i\phi m}$. Hopping around a single plaquette of this lattice imprints a phase $\phi$, analogous to the Aharanov-Bohm phase, emulating a magnetic flux $\Phi/\Phi_0=\phi/2\pi$. We implemented this 2-D lattice by combining a 1-D optical lattice defining sites along an extended direction $\ex$, with atomic spin states forming lattice sites along a narrow, synthetic\cite{Celi2014,Stuhl2015,Mancini2015} direction $\es$.

	This system exhibits a Hall effect, where a longitudinal force $F_{\parallel}$ -- analogous to the electric force $e E_{\parallel}$ in electonic systems -- drives a transverse `Hall' current density $j_{\perp} = \sigma_{\rm{H}} E_{\parallel}$ for non-zero $\Phi/\Phi_0$. A longitudinal force $F_x$ would drive a change in the dimensionless crystal momentum $\hbar\Delta q_x/\hbar G$  and a transverse displacement $\Delta m$, giving a dimensionless Hall conductivity  $N G\Delta m/ \Delta q_x=\sigma_{\rm{H}} R_{\rm{K}}={\tilde\sigma}_{\rm{H}}$, where $G$ is the reciprocal lattice constant and $N$ is the number of carriers per plaquette (see Methods).  Starting with Bose-condensed atoms in the lattice's ground state (with transverse density shown in Fig. \ref{fig:laughlinPump}a) we applied a force along $\ex$ and obtained $\Delta m$ from site resolved density distributions\cite{Wang2013} along $\es$ (Fig. \ref{fig:laughlinPump}b). Leveraging the Diophantine equation, we further show that the force required to move the atoms a single lattice site signals the infinite system's Chern number. 
	
	Our quantum Hall ribbons were created with optically trapped $^{87}\rm{Rb}$ Bose-Einstein condensates (BECs) in either the $F=1$ or $2$ ground state hyperfine manifold, creating $3$ or $5$  site-wide ribbons from the $2F+1$ states available in either manifold. We first loaded BECs into a 1-D optical lattice along $\ex$ formed by a retro-reflected $\lambda_L=1064$ nm laser beam. This created a lattice with period $a=\lambda_L/2$ and depth $4.4(1) E_L$, giving tunneling strength $t_x = 0.078(2) E_L$. Here, $E_L=\hbar^2 k_L^2/2m_{\rm{Rb}}$ is the single photon recoil energy; $\hbar k_L=2\pi\hbar/\lambda_L$ is the single photon recoil momentum; and $m_{\rm{Rb}}$ is the atomic mass. We induced tunneling along $\es$ with strength $t_s = 6.9(4) t_x$ with either a spatially uniform rf magnetic field or two-photon Raman transitions. The rf-induced tunneling imparted no phase, giving $\phi/2\pi = 0$. In contrast the Raman coupling, formed by a pair of counter propagating laser beams with wavelength $\lambda_R=790$ nm, imparted a phase factor of $\exp{(-2ik_Rx)}$. Here, $\hbar k_R=2\pi\hbar/\lambda_R$ is the Raman recoil momentum. Comparing with Eq. \ref{eqn:Hamiltonian}, this gives $\phi/2\pi\approx4/3$. We then applied a force by shifting the center of the confining potential along $\ex$, effectively applying a linear potential. Using time-of-flight (TOF) techniques\cite{Stuhl2015}, we measured hybrid momentum/position density distributions $n(k_x,m)$, a function of momentum along $\ex$ and position along $\es$, as seen in Fig. \ref{fig:laughlinPump}c-d. 
	
\begin{figure}
\includegraphics{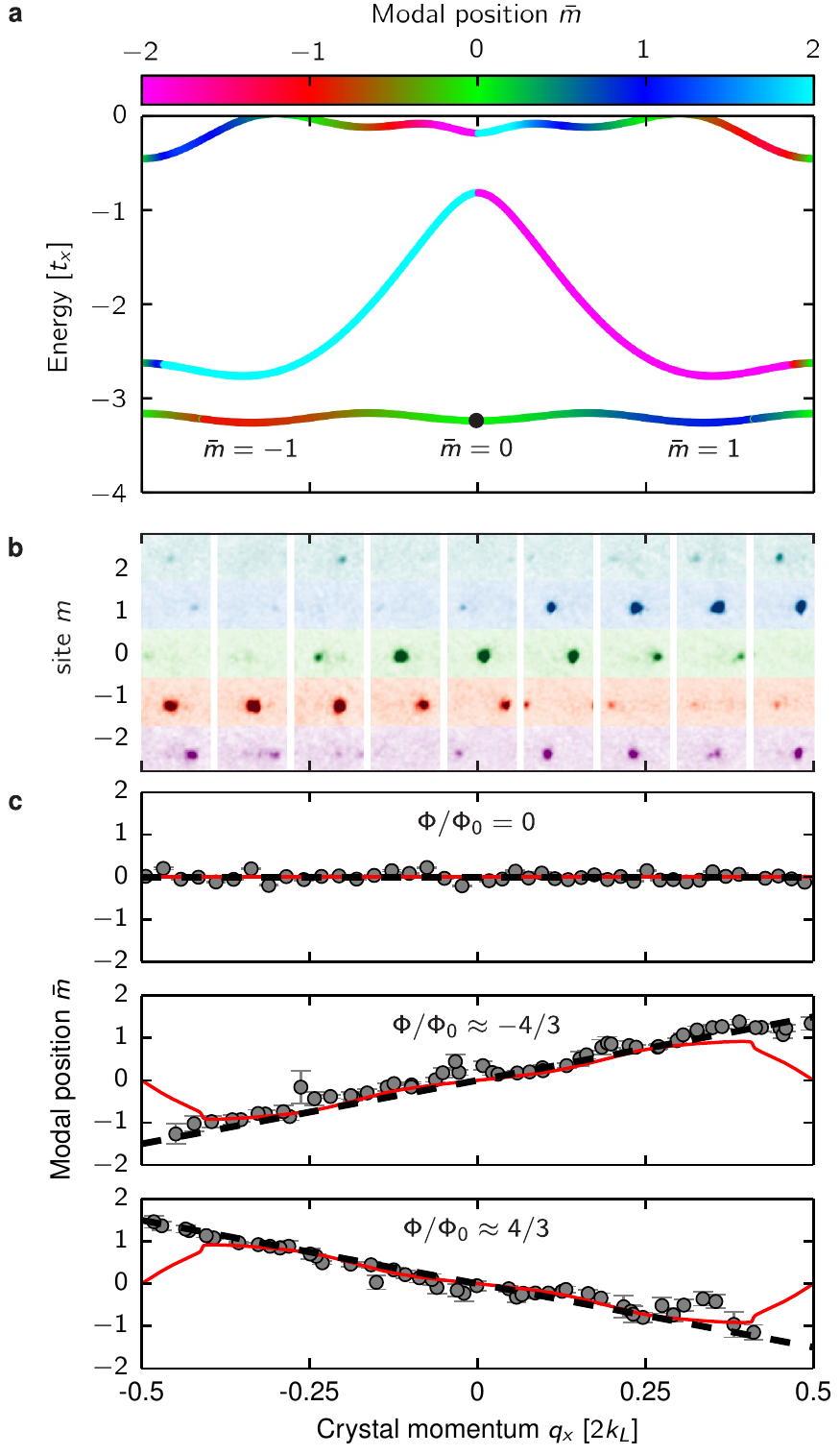}
\caption{Hall displacement in a 5-site wide ribbon.  \textbf{a}. Band structure computed for a $4.4 E_L$ deep 1-D lattice ($\lambda_L=$ 1064 nm), $0.5 E_L$ Raman coupling strength ($\lambda_R = $ 790 nm), and quadratic Zeeman shift $\epsilon=0.02 E_L$, giving $\Phi/\Phi_0 \approx 4/3$, $t_x = 0.078 E_L$, $t_s=6.4 t_x$. The color indicates modal position $\bar{m}$. The black dot indicates the initial loading parameters.  \textbf{b}. TOF absorption images $n(k_x,m)$ for varying longitudinal crystal momenta $q_x$.  \textbf{c}. Transverse displacement. Modal position $\bar{m}$ is plotted as a function of $q_x$ for $\Phi/\Phi_0\approx0,-4/3,4/3$ (top, middle, and bottom respectively). Gray circles depict the measurements; black dashed lines are the prediction of our simple $\tilde{\sigma}_\textrm{H}$ and red curves are the expectation from the band structure of our thin ribbon. As discussed in the methods, the $\Phi/\Phi_0=0$ data was compensated to account for non-adiabaticity in the loading procedure. }
\label{fig:magnetization}
\end{figure}

\begin{figure}
\includegraphics{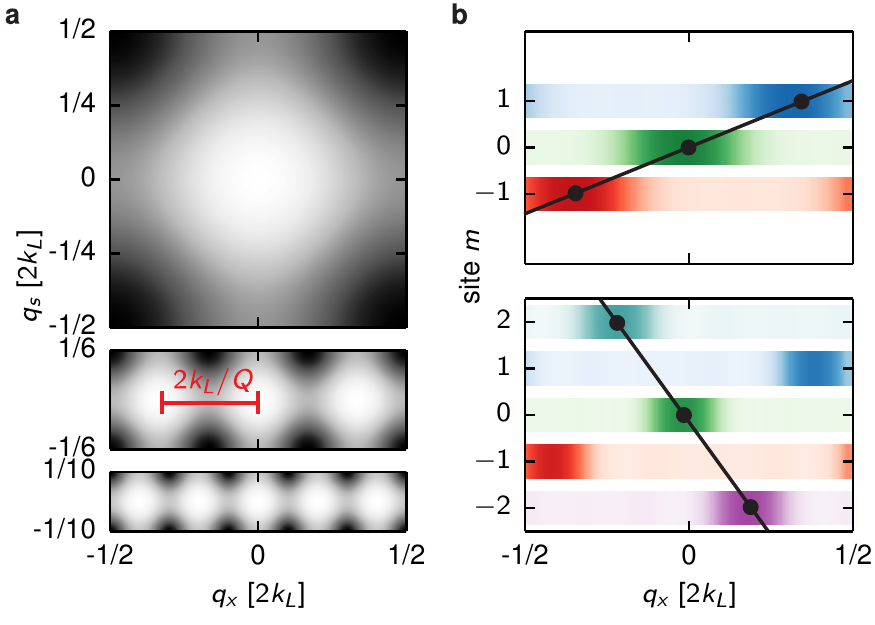}
\caption{Chern number from the Diophantine equation.  \textbf{a}. Lowest band energy within the Brillouin zone in an extended 2-D system, where $q_x$ and $q_s$ are crystal momenta along $\ex$ and $\es$, respectively. Top. $\Phi/\Phi_0=0$. Middle. $\Phi/\Phi_0=1/3$: Brillouin zone shrinks by a factor of $3$ and becomes 3-fold degenerate, distance between adjacent energy minima spaced by $2k_L/Q$ is labeled. Bottom. $\Phi/\Phi_0=2/5$. \textbf{b}.  Fractional population in each spin state in the lowest band at $q_s=0$. Top. $\Phi/\Phi_0=1/3$. Bottom. $\Phi/\Phi_0=2/5$. A momentum shift along $\ex$ of $2k_{\rm L}/Q$ is accompanied by an integer number of spin flips $C$. A line connecting magnetic states separated by $2k_L/Q$, with slope $C=1$ (top) and $-2$ (bottom), is indicated. }
\label{fig:Diophantine}
\end{figure}
	
	We measured the Hall conductivity beginning with a BEC at $q_x(t=0)=0$ in the lowest band with transverse position $\bar{m}_0=0$ (see Methods).  Fig. \ref{fig:magnetization}a shows the band structure of our system as a function of crystal momentum along $\ex$, with color indicating the modal position along $\es$.  We applied a force $F_x$ for varying times $\Delta t$, directly changing the longitudinal crystal momentum from $0$ to a final $q_x$ and giving a transverse Hall displacement from $0$ to a final $\bar{m}$. Figure \ref{fig:magnetization}b shows a collection of hybrid density distributions, where each column depicts $n(k_x,m)$ for a specific final $q_x$, labelled by the overall horizontal axis. For each column, the change in crystal momentum is marked by the horizontal displacement of the diffraction orders relative to their location in the central $q_x=0$ column. The transverse displacement is visible in the overall shift in density along $m$ as a function of $q_x$, i.e., between columns.  

	Figure \ref{fig:magnetization}c quantifies this Hall effect by plotting $\bar{m}$ as a function of $q_x$ for $\Phi/\Phi_0=0$, $-4/3$, and $4/3$. For zero flux $\Phi/\Phi_0=0$ (Fig. \ref{fig:magnetization}c top), $\bar{m}$ was independent of $q_x$; in contrast, for non-zero flux $\Phi/\Phi_0\approx\pm4/3$  (Fig. \ref{fig:magnetization}c middle, bottom),  $\bar{m}$ depends linearly on $q_x$ with non-zero slope. These linear dependencies evoke our earlier discussion of the Hall conductance $\tilde{\sigma}_{\rm{H}}$, in which we anticipated slopes equal to the Chern number. Linear fits to the data give $\tilde{\sigma}_{\rm{H}}=0.01(1)$, $0.87(3)$, and $-0.85(3)$ for zero, negative and positive flux respectively, showing the expected qualitative behavior. The expected slopes, given by the Chern number, $\sigma_{\rm{H}}=0,\pm1$ are indicated by black dashed lines in Fig. \ref{fig:magnetization}c. 

	The red curves in Fig. \ref{fig:magnetization}c show the expected behavior for our 5-site wide system for adiabatic changes in $q_x$, always within the lowest band (Fig. \ref{fig:magnetization}a), i.e., Bloch oscillations.  This theory predicts a nearly linear slope for small $q_x$ sharply returning to $\bar{m}=0$ at the edges of the Brillouin zone. A linear fit to this theory produces $\tilde{\sigma}_{\rm{H}}\approx0$, $0.6$, and $-0.6$ for zero, negative and positive flux respectively, far from the Chern number. This discrepancy is resolved by recalling that Bloch oscillations require adiabatic motion, and the band gaps at the edge of the Brillouin zone close as the ribbon width grows, making the Bloch oscillation model inapplicable. The departure of the data from the adiabatic theory at the edges of the Brillouin zone indicates a partial break down of adiabaticity was present in our data.
	
	To better identify Chern numbers, we relate  the Diophantine equation (Eqn. \ref{eqn:Diophantine}) to the physical processes present in our system.  Although the Hofstadter Hamiltonian in Eqn. \ref{eqn:Hamiltonian} is only invariant under $m$-translations that are integer multiples of $Q$ , a so-called ``magnetic-displacement'' by $\Delta m=1$ accompanied with a crystal momentum shift $\Delta q_x/2 k_{\rm R} = P/Q$ leaves Eqn. \ref{eqn:Hamiltonian} unchanged.  Together, these symmetry operations give a $Q$-fold reduction of the Brillouin zone along ${\bf e}_s$, and add a $Q$-fold degeneracy, as illustrated in Fig. \ref{fig:Diophantine}a  for $\Phi/\Phi_0=0$, $1/3$, and $2/5$.  Recalling that the Brillouin zone is $2 \hbar k_{\rm L}$ periodic along ${\bf e}_x$, it follows that a displacement by $2 k_{\rm L}/Q$ to the nearest symmetry related state involves an integer $C$ magnetic displacements, shown in Fig. \ref{fig:Diophantine}b for $\Phi/\Phi_0=1/3$ and $2/5$, given by solutions to $2  k_{\rm L} s - 2  k_{\rm R} C = 2  k_{\rm L}/Q$, where $s$ counts the number of times the Brillouin zone was ``wrapped around’’ during the $C$ vertical displacements. Because this is no more than a re-expression of the Diophantine equation, we identify $C$ as the Chern number.  Both $C$ and $s$ directly relate to physical processes.  First, each time the  Brillouin zone is wrapped around — implying a net change of momentum by $2 \hbar k_{\rm L}$ — a pair of photons must be exchanged between the optical lattice laser beams.  Second, each change of $m$ by 1 must be accompanied by a $2 \hbar k_{\rm R}$ recoil kick imparted by the Raman lasers as they change the spin state.  This physical motivation of the Diophantine equation remains broadly applicable even for our narrow lattice, providing an alternate signature of the Chern number.

\begin{figure}
\includegraphics{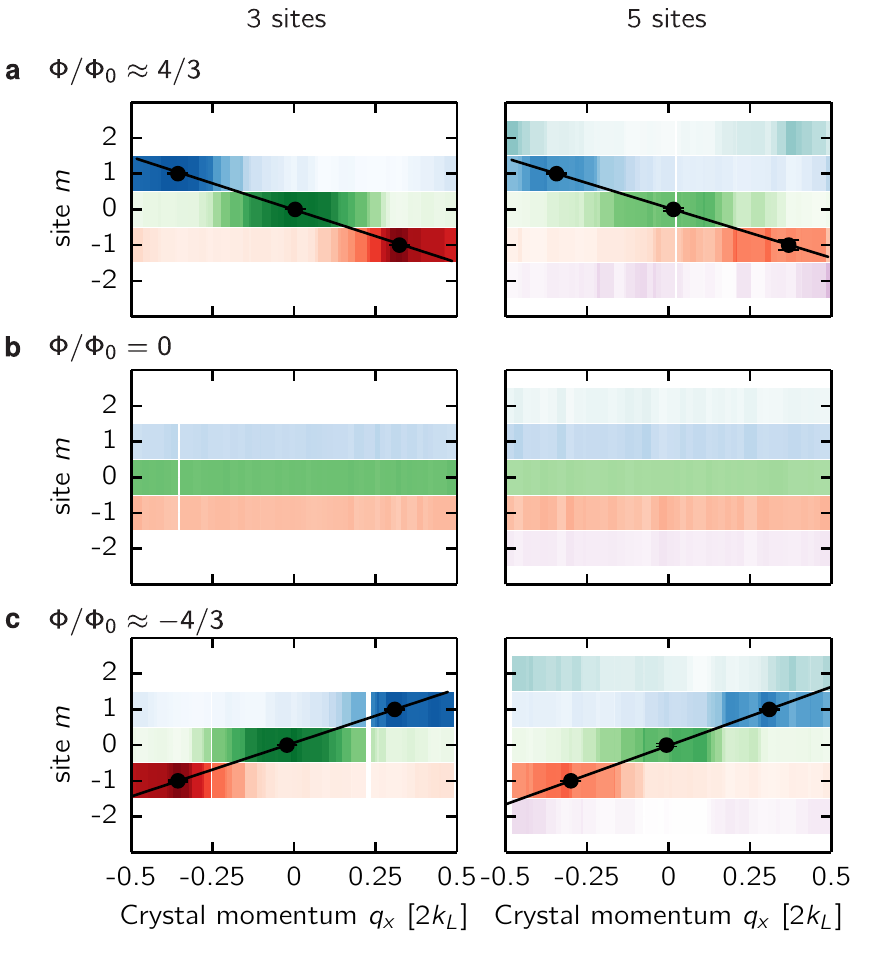}
\caption{Chern number measurement.  Lowest band fractional population measured as a function of crystal momentum in the  $\ex$  and position in the $\es$. Darker color indicates higher fractional population. In the Raman-coupled cases, the points represent the fitted population maxima and the Chern number is extracted from the best fit line to those points.  \textbf{a}. 3-site (left) and 5-site (right) systems with positive flux.  \textbf{b} 3-site (left) and 5-site (right) system with zero flux.  \textbf{c}. 3-site (left) and 5-site (right) systems with negative flux.}
\label{fig:finalData}
\end{figure}

	 Figure \ref{fig:finalData} shows the full evolution of fractional population in each $m$ site as a function of crystal momentum $q_x$ in the lowest band.  The black circles locate the peak of the fractional population in each spin state. We identify the crystal momenta at which the wavefunction was displaced by a single lattice site along $\es$ starting at $q_x=0$, similar to the suggestions in Refs. \cite{Zhang2016,Mugel2017}. We associate the Chern number with the slope of a linear fit through the three peak locations.  For the 3-site wide ribbon, we measured a Chern number of $0.99(4), -0.98(5)$ for negative and positive flux respectively\footnote{Our Chern number extraction scheme fails for the rf case as the fractional populations are flat and there is no peak. We therefore assign a Chern number of $0$ to flat distributions.}, in agreement with the exact theory, which predicts $\pm0.97(1)$.  For the 5-site wide ribbon,  we measured $1.11(2), -0.97(4)$, close to the theoretical prediction of $\pm 1.07(1)$. The deviation from unity results from $\Phi/\Phi_0-4/3\approx0.01$, a non-zero quadratic Zeeman shift, and $t_s>t_x$ allowing hybridization of the edge states\cite{Mugel2017}.
	
	Our direct microscopic observations of topologically driven transverse transport demonstrate the power of combining momentum and site-resolved position measurements. With interactions, these systems are predicted to give rise to complex phase diagrams supporting vortex lattices and charge density waves\cite{Greschner2015,Greschner2016,CalvaneseStrinati2017 }. Realizations of controlled cyclic coupling giving periodic boundary conditions\cite{Celi2014} along $\es$ could elucidate the appearance of edge modes as the coupling between two of the three states is smoothly tuned to zero. In addition, due to the non-trivial topology as well as the low heating afforded by synthetic dimensional systems, a quantum Fermi gas dressed similarly to our system would be a good candidate for realizing fractional Chern insulators\cite{Parameswaran2013}.  
	
\begin{methods}
\subsection{Experiment} We created nearly pure $^{87}\rm{Rb}$ BECs in a crossed optical dipole trap\cite{Stuhl2015} with frequencies $(\omega_x,\omega_y,\omega_z)/2\pi=(27.1(2),58.4(8),94.2(5))\ {\mathrm{Hz}}$. We deliberately used small, low density BECs with $\approx10^3$ atoms to limit unwanted scattering processes in regimes of dynamical instability\cite{Campbell2006}. At various times in the sequence, we used coherent rf and microwave techniques to prepare the hyperfine $\ket{F,m_F}$ state of interest. The 1-D optical lattice was always ramped on linearly in 300 ms. For non-zero $\phi$, we turned on the Raman beams adiabatically in 30 ms after ramping on the lattice. For $\phi=0$ we used adiabatic rapid passage starting in $m_F=-F$ and swept the bias magnetic field in $\approx50$ ms to resonance. We applied forces by spatially displacing the optical dipole beam providing longitudinal confinement (by frequency shifting an acousto-optic modulator), effectively adding a linear contribution to the existent harmonic potential for displacements small compared to the beam waist. 

We define the modal position $\bar{m}$ as the center of a Gaussian fit to the population distribution along $\es$.

\subsection{Rf correction.} In experiments where the tunneling along $\es$ was induced by a uniform rf magnetic field ($\Phi/\Phi_0=0$), our loading procedure had remnant non-adiabaticity that led to temporal oscillations in the fractional populations in different $m$ states at the $40\%$ level. To separate the effects due to this non-adiabaticity from transverse transport, we performed the experiment with identical preparations without applying the longitudinal force. We then used the observed oscillations as a function of time without an applied force as a baseline, and report the difference in fractional populations between that baseline and the cases where the force was applied. 

\subsection{Hall conductivity} The current density can be expressed as $j_{\perp}=n_{2\rm{D}}v_{\perp}e$, where $n_{2\rm{D}}$ is the 2-D charge carrier density, $v_{\perp}$ is the transverse velocity and $e$ is the electron charge. Using $\sigma_{\rm{H}} E_{\parallel}=F_{\parallel}\sigma_{\rm{H}}/e$, and choosing some increment of time $\Delta t$, we have $v_{\perp}=\Delta x_{\perp}/\Delta t$, and $F_{\parallel} = \hbar \Delta q_{\parallel}/\Delta t$, where $q_{\parallel}$ is the crystal momentum along the direction of the force. Re-expressing $n_{2D}$ in number of carriers $N$ per plaquette, defining $\Delta x_{\perp}$ as transverse displacement in units of lattice periods, we obtain $N G\Delta x_{\perp}/ \Delta q_{\parallel}=\sigma_{\rm{H}} R_{\rm{K}}$.

\end{methods}

\bibliographystyle{naturemag}
\bibliography{BlochOscRefs}


\begin{addendum}
 \item [Acknowledgments] This work was partially supported by the Air Force Office of Scientific Research’s Quantum Matter MURI, NIST, and NSF (through the Physics Frontier Center
at the JQI). 
\item [Author Contributions]  D.G. configured the apparatus for this experiment and led the team on all aspects of the measurements. All authors except I.B.S contributed to the data collection effort.  D.G. performed calculations and analyzed data. All authors contributed to writing the manuscript. I.B.S. proposed the experiment concept and wrote this statement.
 \item[Competing Interests] The authors declare that they have no
competing financial interests.
 \item[Correspondence] Correspondence and requests for materials
should be addressed to I.B.S.~(email: ian.spielman@nist.gov).
\end{addendum}

\end{document}